\documentclass[pdflatex,sn-mathphys-num]{sn-jnl}

\usepackage{graphicx}%
\usepackage{multirow}%
\usepackage{amsmath,amssymb,amsfonts}%
\usepackage{amsthm}%
\usepackage{mathrsfs}%
\usepackage[title]{appendix}%
\usepackage{xcolor}%
\usepackage{textcomp}%
\usepackage{manyfoot}%
\usepackage{booktabs}%
\usepackage{algorithm}%
\usepackage{algorithmicx}%
\usepackage{algpseudocode}%
\usepackage{listings}%
 \usepackage{subcaption}
\usepackage{wrapfig}
\usepackage{changes}
\definecolor{feedbackblue}{RGB}{0, 70, 140}

\theoremstyle{thmstyleone}%
\theoremstyle{thmstyletwo}%

\theoremstyle{thmstylethree}%

\begin{document}

\title[Article Title]{Navigating heterogeneous protein landscapes through geometry-aware smoothing}


\author[1]{\fnm{Srinivas}  \sur{Anumasa}}

\author[4]{\fnm{Barath} \sur{Chandran}}

\author[1]{\fnm{Tingting} \sur{Chen}}
\author[1]{\fnm{Nuwaisir} \sur{Mohammad Rahman}}
\author[1]{\fnm{Yingtao} \sur{Zhu}}
\author[1]{\fnm{Rushi} \sur{Shah}}
\author[1]{\fnm{Hongyu} \sur{He}}
\author[1]{\fnm{Peisong} \sur{Zhang}}
\author[1]{\fnm{Yizhen} \sur{Liao}}
\author[1]{\fnm{Yiming} \sur{Tang}}
\author[6]{\fnm{Yong} \sur{Shen}}
\author[5]{\fnm{Tianfan} \sur{Fu}}
\author[7]{\fnm{Rui} \sur{Qing}}
\author[8]{\fnm{Xiao} \sur{Li}}
\author[3]{\fnm{Sebastian} \sur{Maurer-Stroh}}
\author[1,3]{\fnm{Xinyi} \sur{Su}}
\author[2]{\fnm{Zhizhuo} \sur{Zhang}}

\author[1,3]{\fnm{Dianbo} \sur{Liu}}
\affil[1]{\orgname{National University of Singapore}}
\affil[2]{\orgname{GSK.ai,USA}}
\affil[3]{\orgname{A*STAR,Singapore}}
\affil[4]{\orgname{Indian Institute of Technology, Roorkee}}
\affil[5]{\orgname{Nanjing University}}
\affil[6]{\orgname{Xijiao Liverpool University}}
\affil[7]{\orgname{Shanghai Jiaotong University}}
\affil[8]{\orgname{Peking University}}




\abstract{
    The evolutionary fitness landscape of biological molecules is extremely sparse and heterogeneous, with functional sequences forming isolated dense ``islands'' within a vast combinatorial space of largely non-functional variants. Protein sequences, in particular, exemplify this structure, yet most generative artificial intelligence models implicitly assume a homogeneous data distribution. We show that this assumption fundamentally breaks down in heterogeneous biological sequence spaces: fixed global noise levels impose a destructive trade-off, either oversmoothing dense functional clusters or fragmenting sparse regions and producing non-functional hallucinations. To address this limitation, we introduce \emph{Density-Dependent Smoothing} (DDS), a geometry-aware generative framework that adapts stochastic smoothing to the local density of the underlying sequence landscape. By inversely coupling diffusion noise to estimated sequence density, DDS enables gentle refinement in high-density functional regions while promoting controlled exploration across sparse regions. Implemented as a plug-in mechanism for discrete molecular sampling, DDS consistently outperforms state-of-the-art diffusion and autoregressive models across antibody repertoires, therapeutic antibody design, antimicrobial peptide generation and coronavirus antibody design. Together, these results show that fixed global smoothing assumptions fundamentally limit generative modeling in sparse biological sequence spaces, and that geometry-aware smoothing removes this constraint, enabling reliable exploration and design previously unattainable with fixed-noise generative models.

}




\maketitle

\section*{Introduction}\label{sec1}

Molecular design is defined by a fundamental geometric paradox: the ``curse of sparsity'' within an astronomically vast combinatorial space\citep{altman2018curse}. An antibody sequence of length $L$, for example, resides in a manifold of cardinality $20^L$, yet biologically functional proteins occupy only a vanishingly small fraction of this volume \citep{romero2009exploring,russ2020evolution,socolich2005evolutionary,bloom2006protein}. Crucially, these functional sequences are not uniformly distributed\citep{wittmann2021informed,buchholz2018scale}; rather, they form dense, isolated ``islands'' of fitness interspersed within vast, low-density basins of invalid or unstable conformations. Navigating this heterogeneous evolutionary landscape requires a generative model capable of a delicate dualism: it must aggressively explore the sparse ``dark matter'' between islands to discover novel variants, while simultaneously refining dense, high-fitness clusters without distorting their precise functional grammar.

\begin{figure*}[t]
    \centering
    \includegraphics[width=0.95\textwidth]{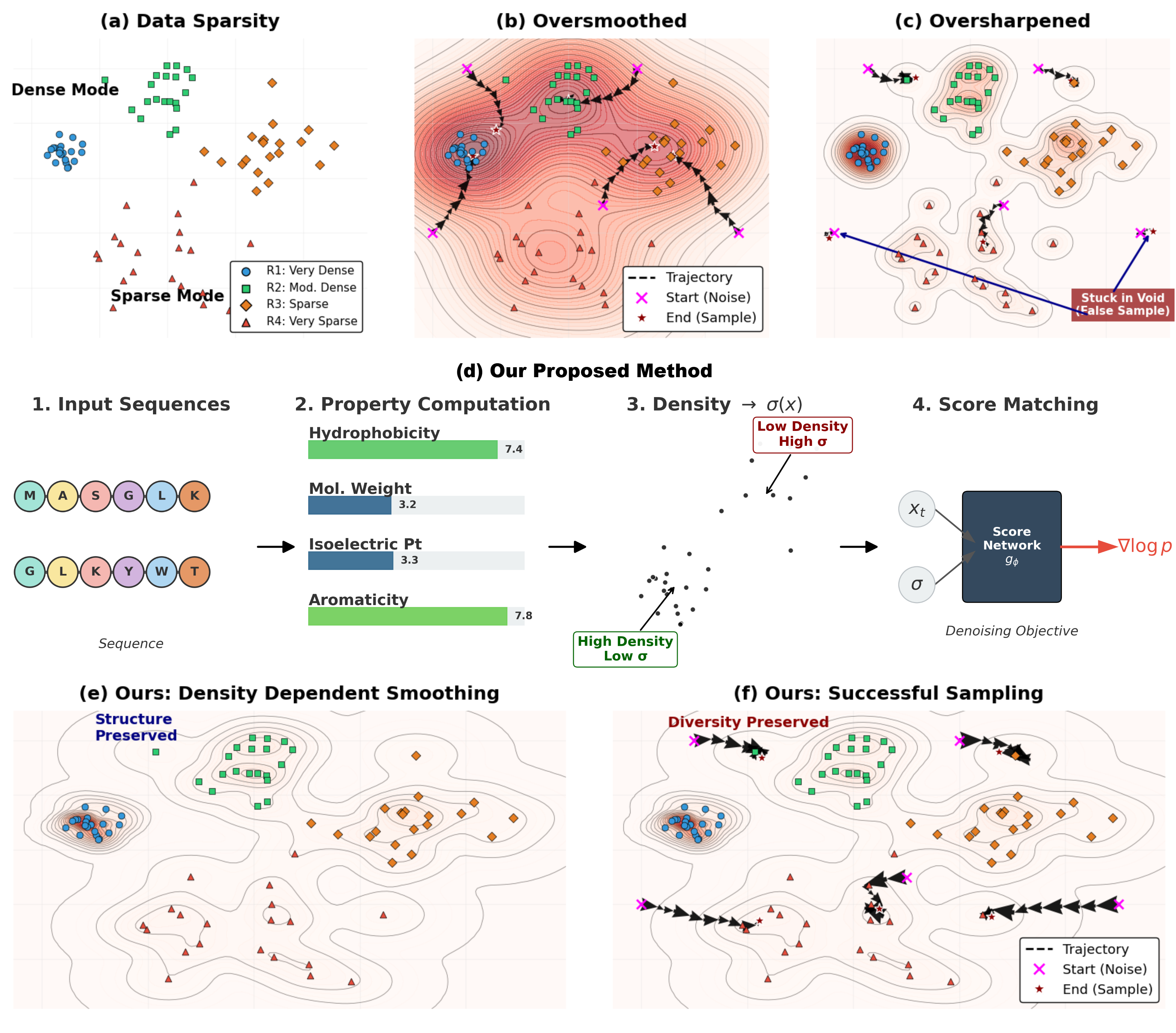}
    \caption{
    \textbf{Fixed noise scales fail to capture the heterogeneous geometry of protein landscapes.} 
    (a) A synthetic dataset illustrating the mixture of dense functional clusters and sparse voids typical of biological data. 
    (b) \textbf{Oversmoothing:} A fixed high noise scale merges distinct functional modes, erasing fine-grained diversity. 
    (c) \textbf{Fragmentation:} A fixed low noise scale creates zero-gradient voids in sparse regions, causing sampling trajectories to stall. 
    (d) \textbf{Density-Dependent Smoothing (DDS):} Our proposed method estimates local density via biophysical properties. 
    (e) The resulting adaptive landscape applies strong smoothing in sparse voids (to bridge gaps) and gentle smoothing in dense clusters (to preserve structure). 
    (f) \textbf{Result:} Sampling trajectories successfully navigate the landscape, recovering diverse modes without hallucinating false positives.
    }
    \label{fig:overview}
\end{figure*}

Current state-of-the-art approaches, particularly discrete diffusion frameworks\citep{frey2023protein,ikram2024gradient}, often fail to reconcile these competing requirements due to a fundamental geometric mismatch. These models typically rely on a fixed global noise scale ($\sigma$) to smooth the data distribution, implicitly assuming a homogeneous landscape. As illustrated in Figure~\ref{fig:overview}, this assumption forces a destructive compromise: a noise level high enough to bridge sparse regions excessively smooths the dense islands, erasing critical structural motifs (oversmoothing, Fig.~\ref{fig:overview}b). Conversely, a noise level low enough to preserve these motifs causes the landscape to fragment in sparse regions, trapping the sampling process in local minima (fragmentation, Fig.~\ref{fig:overview}c). Consequently, existing models must often sacrifice biological plausibility to achieve diversity, or vice versa.

Importantly, this failure is not a consequence of suboptimal hyperparameter tuning. In highly heterogeneous biological sequence spaces, no single global noise scale can simultaneously preserve fine-grained functional structure within dense regions while enabling exploration across sparse regions. As a result, fixed-noise generative models are forced into an unavoidable trade-off between diversity and biological validity. This limitation reflects a fundamental mismatch between global smoothing assumptions and the local geometry of biological fitness landscapes, indicating that reliable generative modeling in this regime requires mechanisms that adapt to local data density rather than relying on a single global scale.

We posit that to navigate this landscape effectively, the ``smoothness'' of the generative process must dynamically match the ``roughness'' of the biological terrain.We propose that diffusion noise should not be treated as a static hyperparameter, but as a geometry-aware control mechanism that adapts to local variation in biological sequence density\citep{song2019generative,kingma2021variational}. To operationalize this insight, we introduce Density-Dependent Smoothing (DDS), a framework that replaces global noise schedules with data-dependent noise scales estimated directly from the training distributionde\citep{de2022riemannian}. By utilizing kernel density estimation to map the local geometry of biochemical feature space, DDS inversely couples the diffusion noise to local sequence density. This allows the generative dynamics to reflect the geometry of the underlying sequence distribution, enabling gentle refinement within high-density functional clusters and broader exploration across sparsely populated regions (Fig.~\ref{fig:overview}d--f).


We validate this approach across therapeutic antibody~\cite{olsen2022observed} and antimicrobial peptide (AMP)~\citep{pirtskhalava2021dbaasp} design tasks, demonstrating that DDS removes the need for a single globally tuned noise scale and enables diffusion-based generative modeling to operate reliably in highly heterogeneous biological sequence spaces.Beyond standard metrics, we utilize ESM-2\citep{lin2023evolutionary} to confirm that DDS generates sequences that are not only diverse but biophysically realizable, exhibiting superior structural folding confidence and predicted binding affinity \citep{lin2023evolutionary}. Crucially, homology analysis ~\citep{altschul1990basic} reveals that DDS transcends the limitations of fixed-noise baselines, which often rely on trivial memorization of training data, to generate genuine evolutionary novelty. These results establish that respecting the heterogeneous geometry of biological data is a prerequisite for reliable, scalable, and biologically grounded \emph{de novo} protein design.

\section*{Result}
We evaluate Density-Dependent Smoothing (DDS) across both controlled synthetic benchmarks and real biological sequence design tasks. We first validate the method on a multi-modal toy dataset with known ground truth to quantify fidelity, mode coverage, and spurious generation. We then assess DDS on large-scale antibody sequence generation using paired repertoires from the Observed Antibody Space (OAS)\citep{olsen2022observed}, followed by therapeutic antibody and antimicrobial peptide benchmarks. For OAS-generated antibodies, we further evaluate biological plausibility using complementary metrics, including structural confidence, language-model likelihood, and evolutionary homology. Together, these results demonstrate the benefits of density-aware smoothing for generative modeling in sparse and heterogeneous protein landscapes. By removing the need for a single globally tuned noise parameter, DDS enables generative models to operate reliably in highly heterogeneous biological sequence spaces, a regime where fixed-noise models systematically fail.

\subsection*{Heterogenous molecule data distribution requires adaptive noise scales.}
A defining characteristic of high-dimensional protein sequence data is its extreme sparsity.  
For an amino acid vocabulary of size $V = 21$ (20 residues plus a padding token) and sequence length $L = 297$,  
the combinatorial space contains $V^L = 21^{297}$ possible configurations.  
Only a vanishingly small subset of this astronomically large space corresponds to biologically viable antibodies,  
meaning that observed sequences occupy isolated, irregular, and highly sparse regions of the underlying distribution.  
Such sparsity poses a fundamental challenge for generative modelling.

\begin{figure}[b]
  \centering
  \begin{subfigure}[b]{0.49\linewidth}
    \centering
    \includegraphics[width=\linewidth]{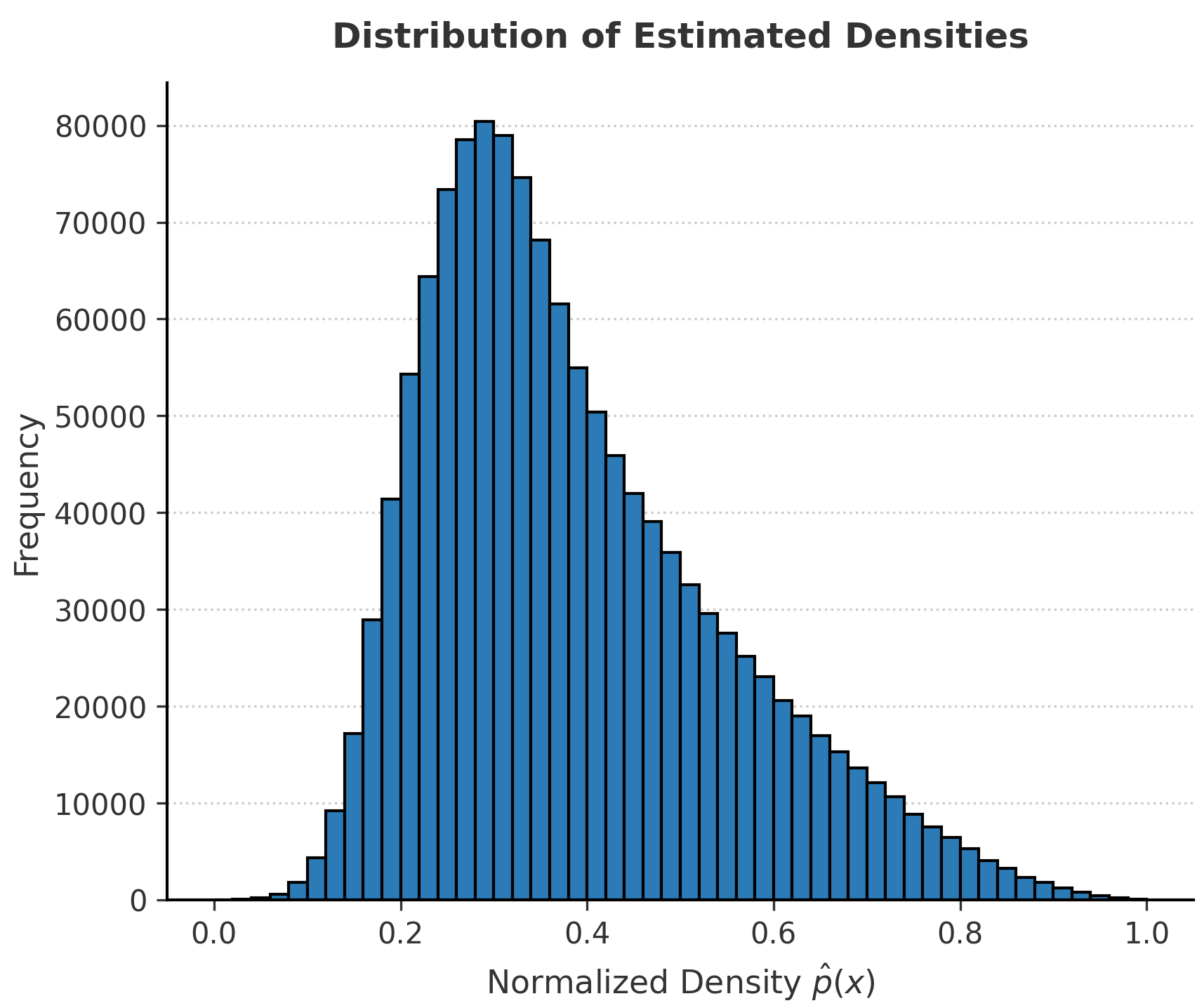}
    \caption{}
  \end{subfigure}
  \begin{subfigure}[b]{0.49\linewidth}
    \centering
    \includegraphics[width=\linewidth]{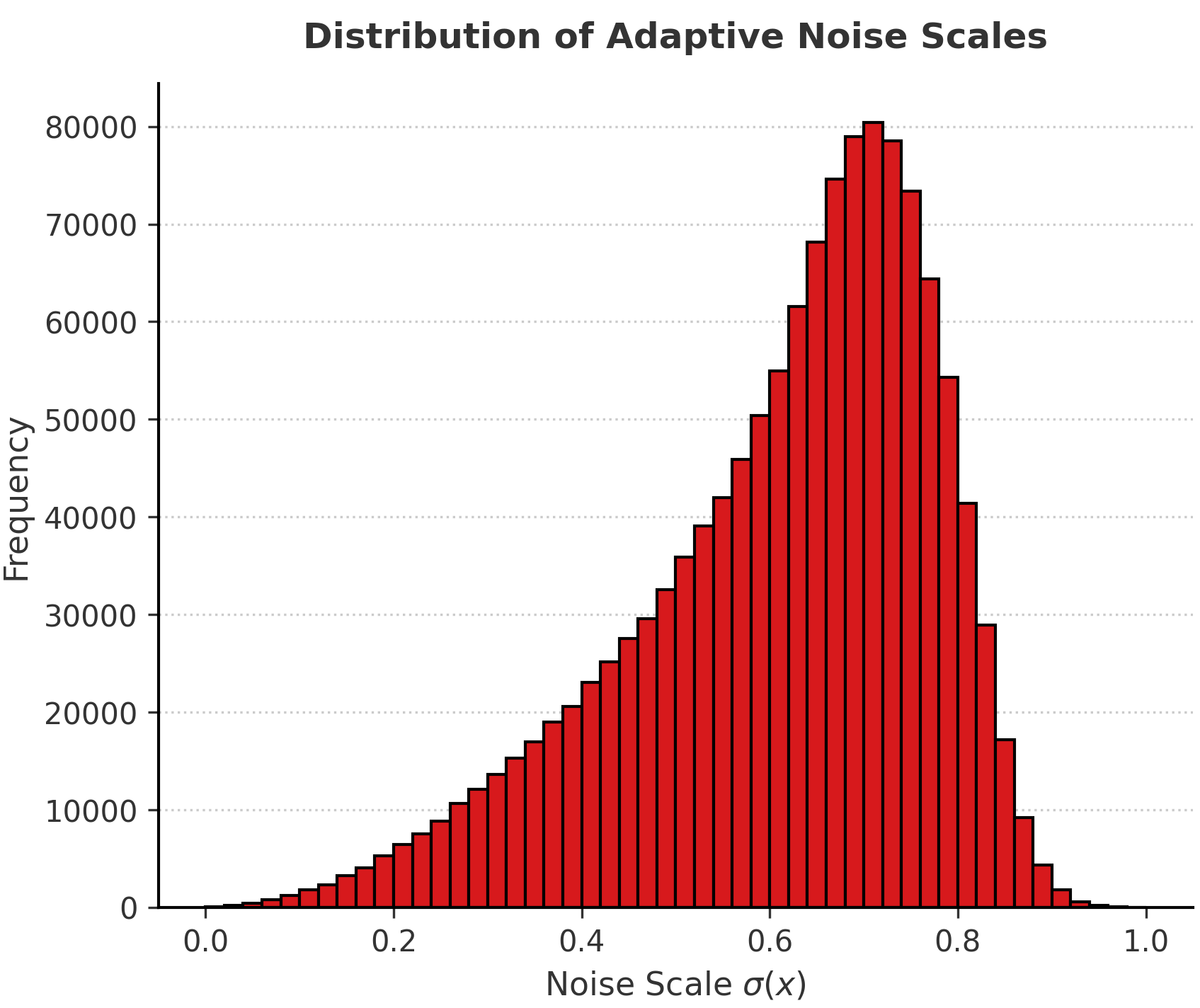}
    \caption{}
  \end{subfigure}
  \caption{Heterogeneity of the antibody sequence space necessitates adaptive smoothing. (a) Distribution of estimated kernel densities $\hat{p}(x)$ for the training dataset. The broad spread of values highlights the varying sparsity of the fitness landscape, where sequences are not uniformly distributed but occupy regions ranging from dense clusters to sparse transitions. (b) Distribution of the corresponding adaptive noise scales $\sigma(x)$. By mapping density to noise inversely, our method assigns a dynamic range of $\sigma$ values, ensuring that regions of varying sparsity receive appropriate levels of smoothing.} 
  \label{fig:density_sigma}
\end{figure}

In frameworks such as dWJS, the choice of the noise level $\sigma$ directly influences how the model navigates this sparse landscape.  
A large $\sigma$ excessively smooths the data manifold, obscuring important structural variations,  
while a small $\sigma$ produces an overly fragmented landscape that can trap the sampler in narrow basins.  
Thus, selecting an appropriate noise level is critical for ensuring both stable training and effective sample generation.

To quantify sparsity in the antibody dataset, we estimate local sample densities using a kernel density estimator (KDE).  
Given training samples $\{x_i\}_{i=1}^{N}$ with feature representations $f(x_i) \in \mathbb{R}^{d}$,  
the KDE estimate at a point $f(x)$ is given by
\begin{equation}
    \hat{p}\big(f(x)\big)
    = \frac{1}{N h^{d}}
      \sum_{i=1}^{N}
      K\!\left(\frac{f(x)-f(x_i)}{h}\right),
    \label{eq:kde}
\end{equation}
where $K(\cdot)$ is a kernel function and $h>0$ denotes the bandwidth.

For the antibody dataset, we project each sequence into a six-dimensional biochemical feature space consisting of  
(a) hydrophobicity, (b) molecular weight, (c) isoelectric point, (d) aromaticity, (e) instability index, and (f) $\beta$-sheet content—  
quantities closely tied to folding, stability, and functional viability.  
Because the full dataset contains more than 1.3 million sequences, we accelerate KDE computation using Random Fourier Features~\citep{rahimi2007random}.  
In contrast, for our synthetic 4D toy dataset, KDE is computed directly in the one-hot embedded space without approximation.

Figure~\ref{fig:density_sigma} (left) shows the distribution of estimated sequence densities, revealing pronounced heterogeneity across the dataset. A small fraction of sequences concentrates within high-density clusters, while the majority occupies varying sparse regions of the biochemical feature space. This observation indicates that sparsity is not uniform but varies substantially across different regions of the sequence landscape. Such heterogeneous sparsity suggests that applying a single, global noise scale during generative modeling is suboptimal, as it fails to accommodate the distinct local structure of dense and sparse regions. This observation motivated as to propose Density-Dependent Smoothing (DDS), which explicitly incorporates local density information to modulate the generative dynamics in a data-aware manner.

Motivated by this observation, we replace the global noise level with a \emph{data-dependent} noise level $\sigma(x)$ that adapts to local density.  
Specifically, we define
\begin{equation}
    \sigma(x) \propto (1- \hat{p}\big(f(x)\big)),
    \label{eq:sigma}
\end{equation}
so that samples in dense regions receive smaller noise levels, while those in sparse regions are assigned larger ones.  
Figure~\ref{fig:density_sigma} (right) shows the resulting distribution of $\sigma$ values, which closely mirrors the empirical sparsity profile of the dataset.

During training, these data-dependent noise levels modulate the denoising objective on a per-sample basis.  
During inference, we sample noise levels $\sigma$ directly from the empirical distribution observed in the training data. As shown in {sampling with a $\sigma$-conditioned score model (Methods), 
each trajectory is initialized and guided using its specific sampled $\sigma$, ensuring that the generation process reflects the heterogeneous sparsity of the protein landscape rather than relying on a single $\sigma$ value.







\subsection*{DDS accurately recovers the underlying distribution of a synthetic 4-dimensional toy dataset.}
To demonstrate that our DDS correctly estimates the underlying distribution, 
we designed a controlled experiment on a synthetic \emph{4-dimensional toy dataset}. 
This setup is meant to mimic key aspects of protein sequences in a simplified setting.  

We consider a vocabulary of size $V=21$ (analogous to the 20 amino acids plus a padding token), 
but restrict the sequence length to $L=4$ instead of 297. 
Thus, the discrete sequence space is
\[
\mathcal{X} \;=\; \{0,1,\dots,20\}^4,
\qquad
|\mathcal{X}| \;=\; V^L \;=\; 21^4.
\]

In analogy to real protein sequences, where only a small fraction of amino-acid sequences 
yield valid proteins due to biochemical constraints, 
we impose an artificial constraint on this toy dataset.  
Specifically, each vocabulary element $v \in \{0,1,\dots,20\}$ is assigned a random 
``hydrophobicity'' score $h(v) \sim \mathrm{Uniform}[-4,4]$. 
For a candidate sequence $x = (x_1, x_2, x_3, x_4) \in \mathcal{X}$, 
we define its hydrophobicity sum as
\[
H(x) \;=\; \sum_{i=1}^4 h(x_i).
\]

A sequence is considered \emph{valid} if and only if it satisfies the threshold criterion
\[
H(x) \;\geq\; \tau,
\]
where $\tau$ is a fixed constant (e.g., $\tau=20$ in our experiments).  

Formally, the set of valid sequences is
\[
\mathcal{X}_{\text{valid}} \;=\; \{\, x \in \mathcal{X} \;:\; H(x) \geq \tau \,\}.
\]

By exhaustively enumerating all possible sequences $x \in \mathcal{X}$ 
and filtering according to the above constraint, 
we obtain a sparse distribution of valid points in 4D space. 
This mirrors the behavior of real protein sequence data, 
where the vast majority of possible sequences are invalid, 
and the observed data lie in sparse, structured regions of the combinatorial space.
\begin{figure}[t]
  \centering
  \begin{subfigure}[b]{0.47\linewidth}
    \centering
    \includegraphics[width=\linewidth]{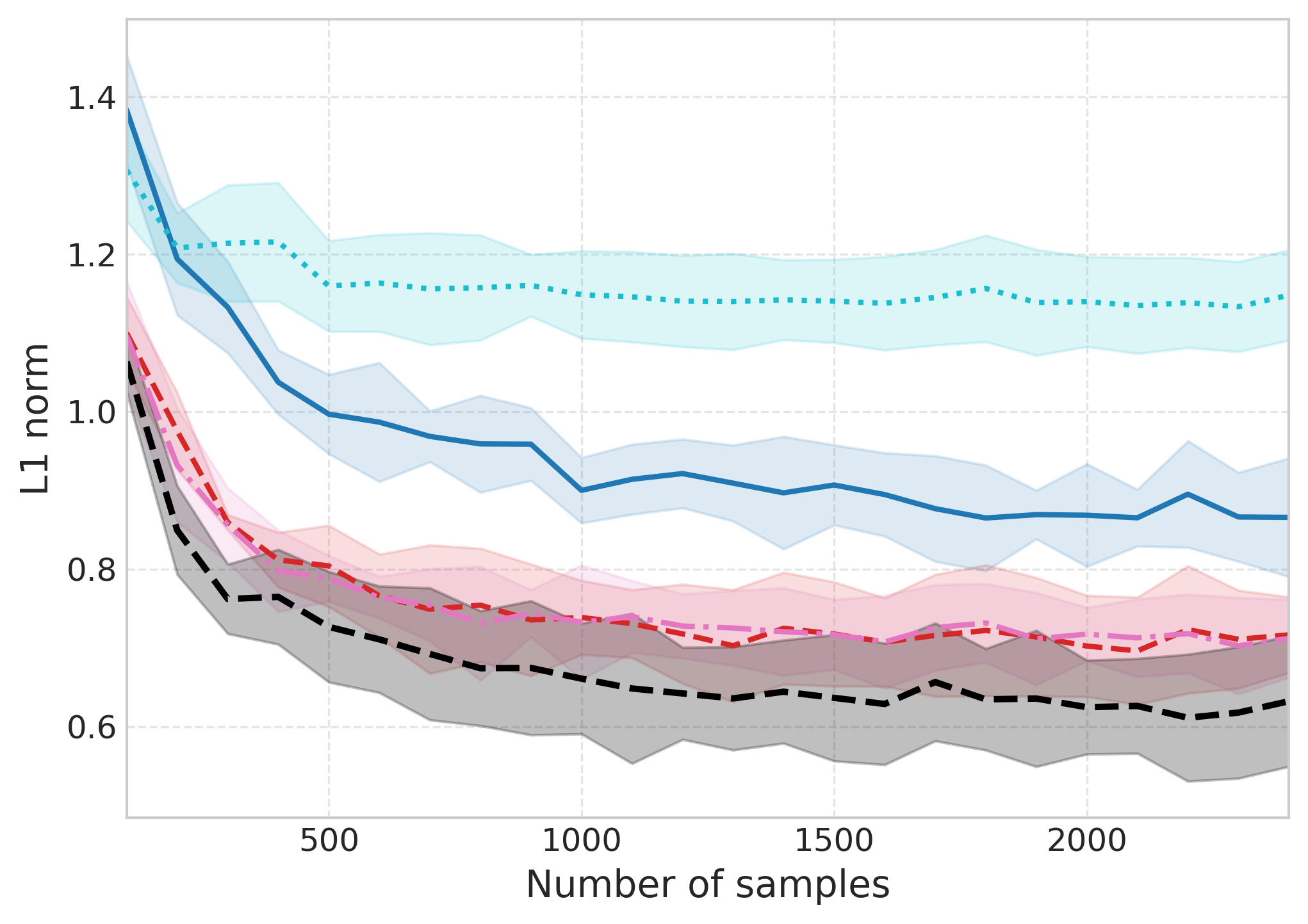}
    \caption{$L_1$ norm}
    \label{fig:l1}
  \end{subfigure}\hfill
  \begin{subfigure}[b]{0.47\linewidth}
    \centering
    \includegraphics[width=\linewidth]{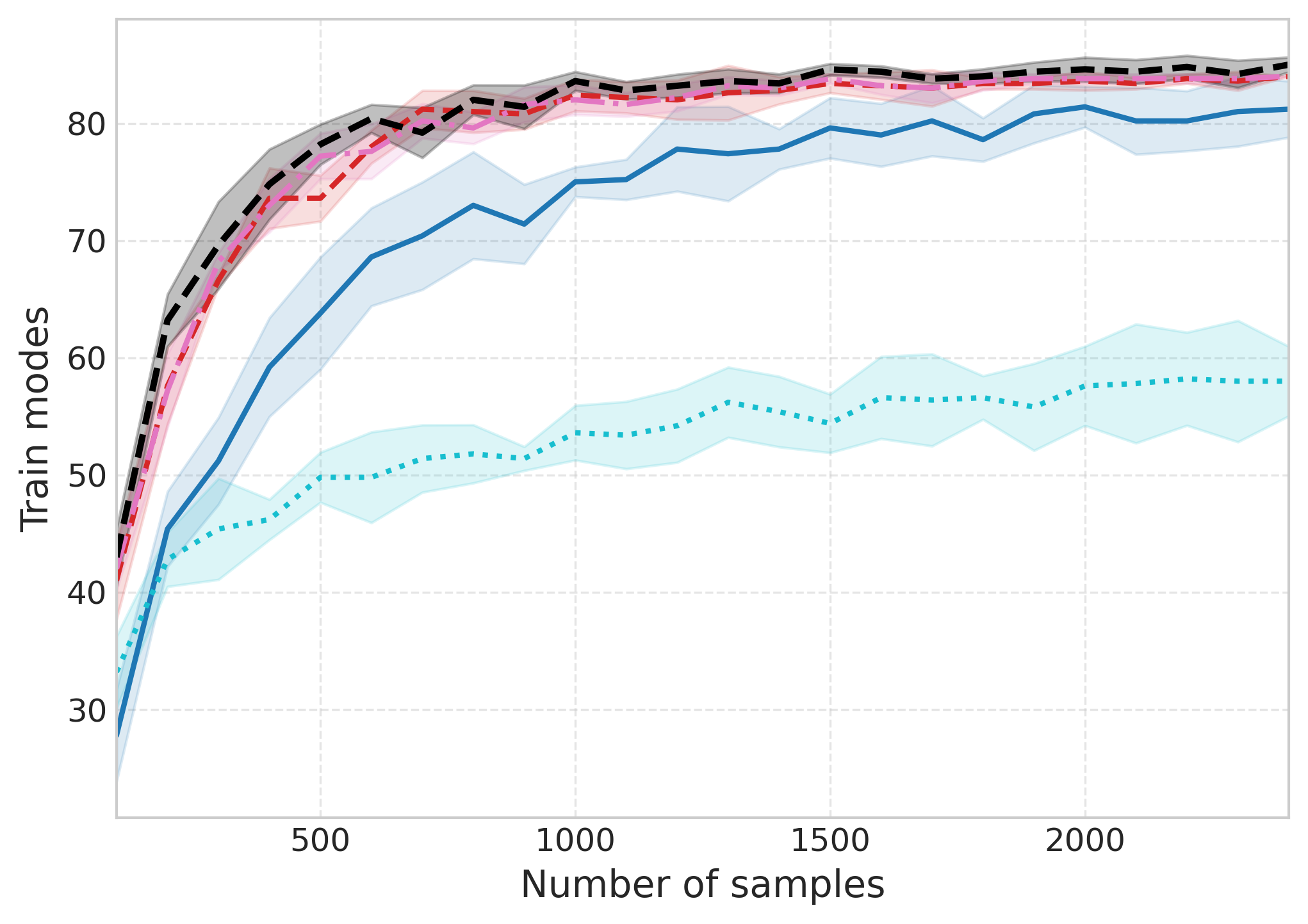}
    \caption{True modes covered}
    \label{fig:true}
  \end{subfigure}\hfill
  \begin{subfigure}[b]{0.47\linewidth}
    \centering
    \includegraphics[width=\linewidth]{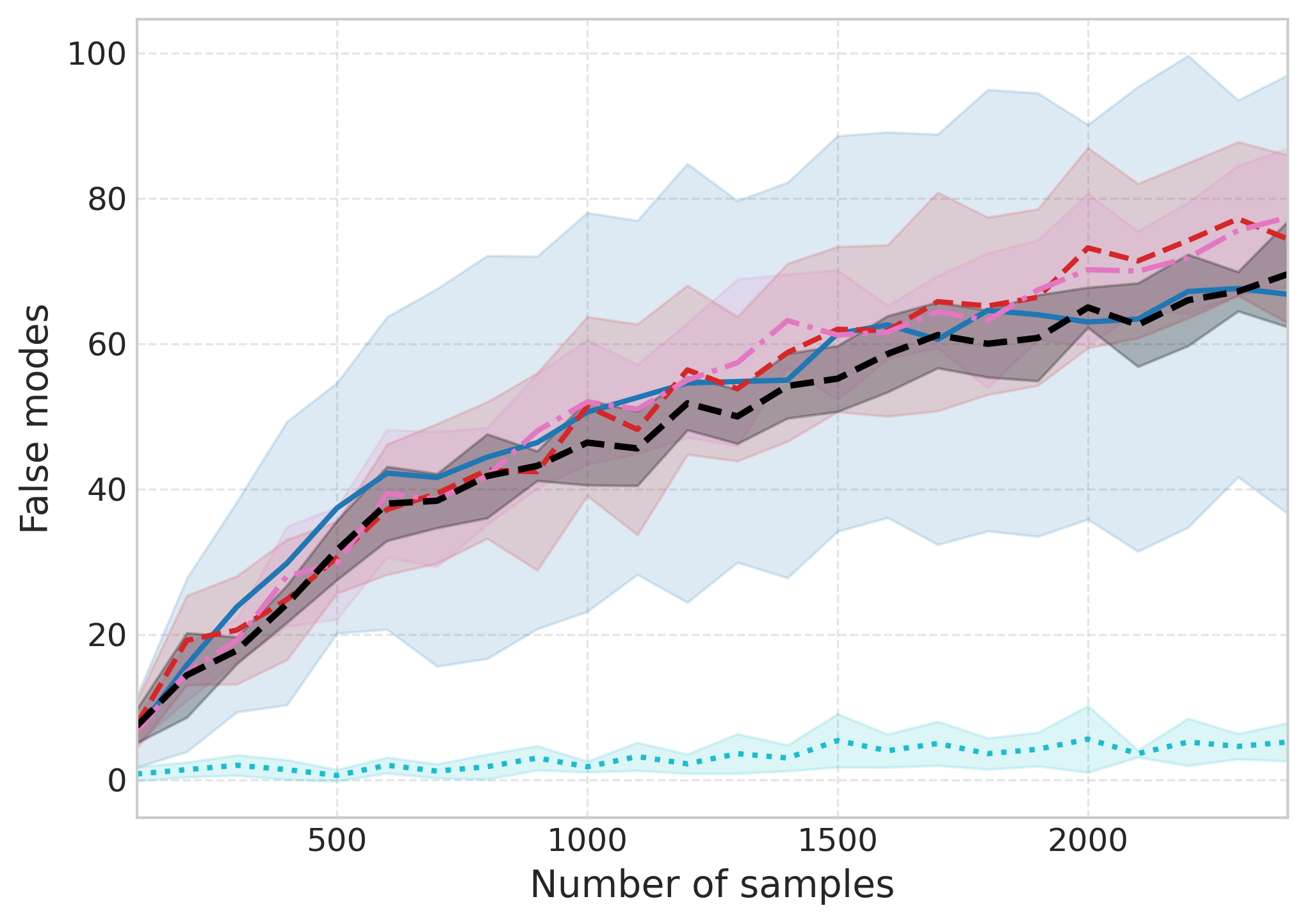}
    \caption{False modes covered}
    \label{fig:false}
  \end{subfigure}\hfill
    \begin{subfigure}[b]{0.47\linewidth}
    \centering
    \includegraphics[width=\linewidth]{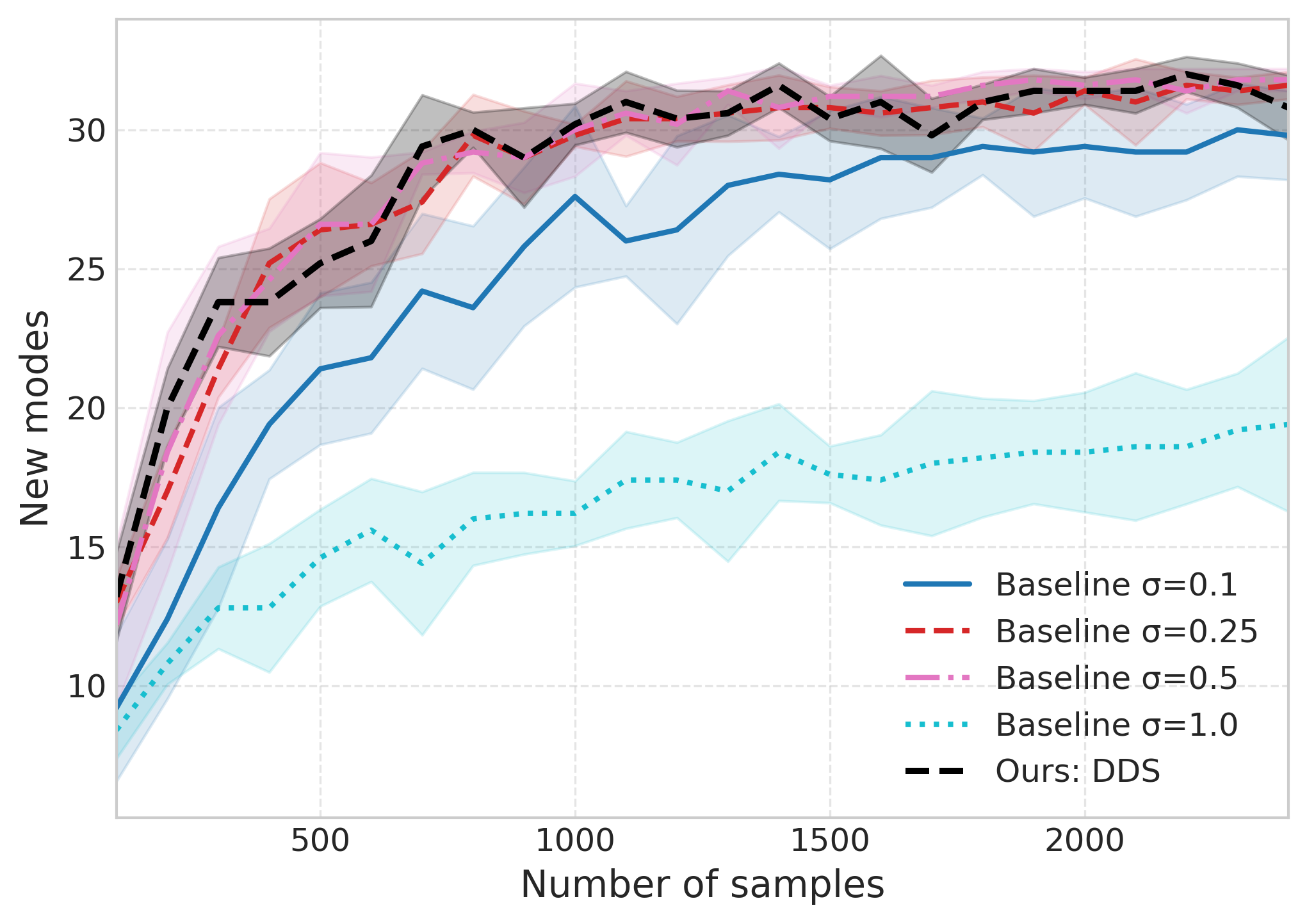}
    \caption{New modes covered}
    \label{fig:false}
  \end{subfigure}

  \caption{Data-dependent smoothing improves fidelity–exploration balance on a synthetic multimodal landscape. (a) \textbf{$L_1$ distance} between the generated and ground-truth distributions as a function of the number of samples. Density-Dependent Smoothing (DDS; black dashed) consistently achieves lower error, indicating improved approximation of the target distribution. (b) \textbf{Number of true training modes recovered}. DDS attains recall comparable to well-tuned low-noise baselines, whereas large-noise models under-cover modes. (c) \textbf{Number of false (spurious) modes generated}. Low-noise baselines increasingly generate false modes with additional sampling, while DDS maintains substantially lower spurious mode counts. (d) \textbf{Number of novel (held-out) modes recovered.} DDS generalizes beyond training modes, recovering unseen modes while preserving fidelity.
Shaded regions indicate mean ± s.d. over $n=5$ independent runs.}
  \label{fig:synthetic_modes}
\end{figure}

We evaluate DDS against standard dWJS baselines with fixed noise scales
$\sigma \in \{0.1, 0.2, 0.25, 0.35, 0.50, 0.75, 1.0\}$. For DDS, we restrict the noise
range to $\sigma \in [0.25, 0.35]$. This choice is guided by empirical observations
from the baseline models, where intermediate noise levels ($\sigma = 0.25$ and
$0.35$) consistently yield better performance than either lower or higher values.
Accordingly, DDS is configured to interpolate within this empirically effective
noise regime.
 
The synthetic dataset consists of 120 ground-truth modes, of which 80\% are used for training, while the remaining 20\% are held out for assessing recovery. 
All models share the same ByteNet backbone and are trained for 500 epochs. 

For evaluation, we generate samples using an increasing number of random samples, with each inference stage producing more samples. 
For every noise configuration, we train five independent models with different random seeds. 
Figure~\ref{fig:synthetic_modes} reports the mean and standard deviation across these runs for four complementary metrics. For clairifty here we only compare the baselines with 4 different $\sigma$ values, in \ref{fig:synthetic_modes_full} we compared with baselines with all the 7 different $\sigma$ values.

\paragraph{Adaptive smoothing reduces L1 error across all sample budgets}
As shown in Figure~\ref{fig:synthetic_modes}(a), the $L_1$ distance decreases for all models as more samples are generated, indicating improved approximation of the ground-truth distribution. 
Across all sample budgets, DDS consistently achieves the lowest $L_1$ error among all baselines. 
This demonstrates that data-dependent smoothing produces sigma values that better reflect local data density, enabling the model to recover the true distribution more faithfully than fixed-$\sigma$ baselines.

\paragraph{DDS balances mode recovery and generalization while avoiding over-smoothing.}
Figures~\ref{fig:synthetic_modes}(b) and (d) show the number of \emph{true modes recovered} during sampling and the number of \emph{held-out or new modes discovered}. 
DDS and the baselines with $\sigma \in \{0.25\}$ recover the majority of ground-truth modes as the sampling budget increases. 
In contrast, baselines with larger fixed $\sigma$ values (e.g., $0.5$ and $1.0$) recover significantly fewer modes, indicating that over-smoothing suppresses exploration of the underlying mode structure. On the mode-discovery task, DDS matches closely with the best fixed-$\sigma$ baselines, successfully identifying held-out modes that were never shown during training. 
This illustrates that our data-dependent smoothing strategy allows controlled exploration that generalizes beyond the training support.
\paragraph{DDS balances exploration and correctness, minimizing false mode generation.}
Figure~\ref{fig:synthetic_modes}(c) highlights the trade-off between exploration and correctness. 
While aggressive exploration may increase diversity, it often leads to \emph{false modes}, samples that do not correspond to any ground-truth structure. 
DDS achieves a favorable balance: it effectively recovers true modes and discovers new modes (Figures~\ref{fig:synthetic_modes}(b,d)), while generating substantially fewer false modes than aggressively sharpened baselines.  In contrast, models with small fixed $\sigma$ values tend to over-explore, producing numerous spurious samples outside the true distribution. 
DDS avoids this failure mode, remaining close to the correct data manifold while still exploring meaningfully beyond the training distribution.

\subsection*{DDS produces sequences that are both novel and biophysically plausible.}
We evaluate our proposed DDS model on variable-length antibody sequences from the Observed Antibody Space (OAS) database~\cite{olsen2022observed}. Each antibody sequence is represented as $x = (x_1, \ldots, x_d)$, where $x_l \in \{1, \ldots, 20\}$ denotes the amino acid (AA) type at position $l$. All sequences are aligned using the AHo numbering scheme and subsequently one-hot encoded. Because antibody sequences have variable lengths, we append a gap token to obtain a fixed-length representation: heavy-chain sequences are padded to length $149$, and light-chain sequences to length $148$, where each position may take one of $21$ possible symbols (20 AAs plus a gap token). After alignment and encoding, each antibody is represented as a binary vector of dimension $(149 + 148) \times 21$. For POAS we particularly focus on discovery of new samples, initilised from a random noise, guided by a trained $\sigma$ conditioned score model we generate new samples. 

We compare our model against a diverse set of baseline methods spanning diffusion-based approaches and autoregressive language models. Our primary baseline is dWJS~\citep{frey2023protein}, on top of which our method is built. In addition, we evaluate against: the latent sequence diffusion model SeqVDM, a discrete extension of variational diffusion~\citep{kingma2021variational}; the score-based model DEEN, which employs an energy parameterization~\citep{saremi2018deep}; IgLM, a transformer-based language model tailored for antibody design~\citep{shuai2021generative}; ESM2~\citep{lin2023evolutionary}; and a large pre-trained language model (GPT-3.5).

\begin{table}[h!]
\centering
\caption{Comparison of model performance with baselines on the antibody sequence design.} 
\label{tab:protein_results}
\begin{tabular}{lccccc}
\toprule
\textbf{Model / $\sigma$} & \textbf{U $\uparrow$} & \textbf{ID $\uparrow$} & \textbf{ED $\uparrow$} & \textbf{DCS $\uparrow$} & \textbf{WD $\downarrow$} \\
\midrule
\textbf{dWJS ($\sigma$=1.0)} & 1.00 & 95.59 ± 24.58 & 34.38 ± 19.60 & 0.45 ± 0.30 & 0.072 ± 0.075 \\
\textbf{dWJS ($\sigma$=0.5)} & 1.00 & \textbf{119.29 ± 20.94} & 61.71 ± 18.88 & 0.25 ± 0.28 & 0.068 ± 0.046 \\

\textbf{DDS}         & 1.00 & 106.92 ± 22.33 & 50.71 ± 17.57 & 0.45 ± 0.30 & \textbf{0.057 ± 0.024} \\
\textbf{SeqVDM} \citep{kingma2021variational}              & 1.00 & 57.40 & 60.00 & 0.40 & 0.062 \\
\textbf{DEEN} \citep{saremi2018deep}               & 0.99 & 42.70 & 50.90 & 0.41 & 0.087 \\
\textbf{GPT-3.5}             & 0.66 & 46.10 & 55.40 & 0.23 & 0.140 \\
\textbf{IgLM} \citep{shuai2021generative}               & 1.00 & 34.60 & 48.60 & \textbf{0.53} & 0.080 \\
\textbf{ESM2} \citep{lin2023evolutionary}                & 1.00 & 77.56 & \textbf{70.99} & 0.06 & 0.150 \\
\bottomrule
\end{tabular}
\footnotesize{         $^1$We reproduced the dWJS results, whereas the performance numbers for the remaining baselines were obtained from the results reported in the dWJS study, where they did not provide std values\citep{frey2023protein}.}
\end{table}

We trained our proposed DDS model and the baseline dWJS using the same experimental setup described in \citet{frey2023protein}.
Both models were trained with a ByteNet backbone for 40 epochs, using identical architectural choices, learning rate, and weight decay settings to ensure a fair comparison. For the baseline dWJS, we evaluated two fixed noise-scale hyperparameters ($\sigma \in {0.5, 1.0}$). In contrast, our DDS model employs a data-dependent smoothing mechanism in which per-sample $\sigma$ values are estimated via KDE from the training data and then normalized and rescaled to fall within the range $[0.4, 0.6]$. To assess the quality of generated antibody sequences, we evaluate a diverse set of metrics that capture uniqueness, diversity, novelty, distributional similarity, such as, Uniqueness (U), Intra-Diversity (ID), Edit Distance (ED), Distributional Consistency Score (DCS), and Wasserstein Distance (WD) (methods).

Table~\ref{tab:protein_results} summarizes the performance of our approach in comparison with dWJS and the additional baselines. A strong generative model should balance diversity and quality: the generated sequences must be sufficiently different from each other and different than the samples used to train the model, while still remaining close to the true underlying distribution. From Table~\ref{tab:protein_results}, we observe that our DDS model achieves this balance more effectively than the baselines. In particular, DDS maintains competitive diversity (ID and ED) while producing higher-quality samples, as reflected in the DCS and WD metrics.

In contrast, dWJS with a large fixed noise scale ($\sigma = 0.5$) exhibits high diversity but at the cost of sample quality—evidenced by a significantly worse WD score—indicating that its uncontrolled exploration pushes the model toward unrealistic regions of sequence space. Although dWJS($\sigma=0.5$) performs well under the ED metric , suggesting it generates sequences distinct from the validation set, many of these sequences are low-quality or dysfunctional.

Overall, DDS produces sequences that are both novel and biophysically plausible, yielding a superior trade-off between exploration and realism. This trend aligns with our synthetic experiments, where DDS successfully recovered more true modes while avoiding false ones. The consistency across all metrics highlights the advantage of using data-dependent smoothing over fixed-noise generative strategies.

Prior work \citep{frey2023protein} suggests that values above $0.3$ are sufficient to generate samples that are viable in wet-lab settings. 
Thus, DDS provides an effective balance between quality and distributional coverage.

\subsection*{DDS maintains higher diversity while preserving biological plausibility in CDR mutants generation.}
Next, we evaluate our method on the task of generating CDR mutants using the hu4D5 antibody mutant dataset \citep{mason2021optimization}. After de-duplication and removal of samples with conflicting labels, the dataset consists of approximately 9k binding and 25k non-binding hu4D5 CDR mutants, each containing up to 10 mutations. To evaluate the quality of the generated sequences, we trained a binary classifier achieving 85\% accuracy on an i.i.d. validation set. All generative models were trained on the full dataset comprising 34k hu4D5 CDR H3 mutants. Following the evaluation protocol introduced in GFlowNets~\citep{jain2022biological}, we generate 500 sequences, from which the top 100 ranked according to the classifier’s predicted binding probability are retained for analysis. Table~\ref{tab:her2_results} reports the comparative performance across methods. We want a generative model to balance both the diversity and quality of the samples. 

GFlowNets exhibit high intra-diversity (ID), yet many of their generated sequences receive low predicted binding probability, indicating that they may fall outside the distribution of true binders. In contrast, gg-dWJS~\citep{ikram2024gradient} attains strong binding probability ($P_{\text{bind}}$) but produces less diverse samples. Our DDS model strikes a more desirable balance, it maintains higher diversity while preserving biological plausibility, as reflected in the harmonic mean (HM) metric, demonstrating its effectiveness in generating sequences that are both novel and functionally promising.
\begin{table}[h!]
\centering
\caption{Comparison of binding probability and diversity metrics across models. }
\label{tab:her2_results}
\begin{tabular}{lcccc}
\toprule
\textbf{Model} & \textbf{$P_{\text{bind}}(\uparrow)$} & \textbf{ID}($\uparrow$) & \textbf{ED}($\uparrow$) & \textbf{HM}($\uparrow$) \\
\midrule

\textbf{dWJS} ($\sigma=1.0$) & 0.96 ± 0.01 & 6.15 ± 1.37 & 2.66 ± 0.68 & 0.41 ± 0.08 \\
\textbf{dWJS} ($\sigma=0.5$) & 0.95 ± 0.02 & 6.60 ± 1.28 & 2.68 ± 0.56 & 0.41 ± 0.07 \\
\textbf{DDS} & 0.93 ± 0.03 & 6.82 ± 1.32 & 2.82 ± 0.63 & \textbf{0.43 ± 0.07} \\
\textbf{gg-dWJS} & \textbf{0.98 ± 0.01} & 4.58 ± 1.17 & 2.24 ±0.58 & 0.36±0.08 \\
\textbf{GPT-4o} & 0.58 ±0.36 & 7.42±1.57 & \textbf{3.29±0.96} & 0.34±0.16 \\
\textbf{SeqVDM} & 0.36 ±0.26 & 7.48±1.21 & 3.25±0.64 & 0.24±0.16 \\
\textbf{GFlowNets}  & 0.32 & \textbf{9.16} & - & - \\
\bottomrule
\end{tabular}
\footnotesize{         $^1$We reproduced the dWJS results, whereas the performance numbers for the remaining baselines were obtained from the results reported in the gg-dWJS study and their github repository\citep{ikram2024gradient}.}
\end{table}

\subsection*{DDS achieves a favourable balance between sequence quality and diversity in Anti-microbial peptide (AMP) generation.}
This experiment aims to generate peptides with anti-microbial properties. We construct our dataset from AMPs and non-AMPs in the DBAASP database~\citep{pirtskhalava2021dbaasp}. Following the protocol of Jain et al.~\citep{jain2022biological}, we select peptides of length 12--60 that target Gram-positive bacteria, yielding 6,438 AMPs and 9,222 non-AMPs.

We split the dataset into two subsets, D1 and D2, following~\citep{angermueller2019model}, D1 is provided to the generative model, while D2 is used exclusively to train an oracle classifier for evaluation. The split enforces the same group-disjointness rule as in Jain et al., where for any peptide $x \in D1$, there is no peptide $x' \in D2$ belonging to the same group, and vice versa. This results in 3,219 AMPs and 4,611 non-AMPs in D1. We train the oracle using a classifier that achieves 91\% accuracy. Following~\citep{jain2022biological},  we evaluate the models on 3 different metrics, along with Harmonic-mean, which measures both quality and diversity. 

\textbf{Harmonic Mean (HM)\citep{vishakh2025measuring}.} 
The HM metric captures the trade-off between \emph{quality} and \emph{diversity} of the generated samples. 
For each generated sequence, the quality score is obtained from the trained classifier, while the diversity score is computed using the normalized edit distance (edit distance divided by the sequence length; for AMP sequences this length is $60$). 
The harmonic mean of these two quantities encourages models to generate samples that are both high-quality and diverse, rather than excelling in only one dimension.

\begin{table}[h!]
\centering
\caption{Comparison of model performance between ours and baselines on the AMP dataset.}
\label{tab:amp_results}
\begin{tabular}{lcccc}
\toprule
\textbf{Method} & \textbf{H-Mean($\uparrow$)} & \textbf{Performance ($\uparrow$)} & \textbf{ID ($\uparrow$)} & \textbf{ED ($\uparrow$)} \\
\midrule
\textbf{DDS} & \textbf{0.73 ± 0.08} & 0.92 ± 0.10 & 45.93 ± 5.40 & 36.80 ± 6.41 \\
\textbf{dWJS ($\sigma$=0.5)} & 0.65 ±0.15 & 0.70 ±0.27 & \textbf{48.83  ± 3.91 } & \textbf{40.89  ± 4.79} \\
\textbf{dWJS ($\sigma$=1.0)} & 0.36 ±0.13 & \textbf{0.99 ±0.01} & 21.21  ± 6.54 & 13.87  ± 5.97\\
\textbf{gg-dWJS} & -- & 0.98 & 25.78 & 15.02 \\
\textbf{GFlowNet-AL} & -- & 0.93 & 22.34 & 28.44 \\
\textbf{DynaPPO} & -- & 0.94 & 12.12 & 9.31 \\
\textbf{COMs} & -- & 0.76 & 19.38 & 26.47 \\
\textbf{GFlowNet} & -- & 0.87 & 11.32 & 15.72 \\
\bottomrule
\end{tabular}
\footnotesize{         $^1$We reproduced the dWJS results, whereas the performance numbers for the remaining baselines were obtained from the results reported in the gg-dWJS study\citep{ikram2024gradient}.}
\end{table}

Table~\ref{tab:amp_results} summarises the performance of the different models across multiple evaluation metrics. The metric \textit{Performance} measures the average probability assigned by the trained classifier that a generated peptide is an AMP. Following the evaluation protocol introduced in GFlowNets~\citep{jain2022biological}, each model generates 500 sequences, from which the top 100, ranked by the classifier’s predicted AMP probability, are retained for analysis. 

Under the ID and ED metrics, dWJS with $\sigma = 0.5$ performs relatively well; however, its corresponding \textit{Performance} score indicates that many of its generated samples are unlikely to be AMPs according to the classifier. In contrast, gg-DWJS achieves a strong \textit{Performance} value but produces less diverse samples.

Our model achieves a favourable balance between sequence quality and diversity, as reflected by the harmonic mean (HM) metric. Notably, all top-100 samples generated by our method are classified as AMPs, but with predicted probabilities ranging from 0.85 to 0.95, yielding an average value of 0.92. This spread highlights the model’s ability to generate a diverse yet biologically plausible set of AMP candidates.

\subsection*{Density-based $\sigma$ estimation is critical for quality and diversity of generated samples.}
To verify whether the skewed $\sigma$ distribution obtained from density estimation truly contributes to the performance of our model, we conducted an experiment in which the $\sigma$ values were \emph{uniformly} ($\sigma= 0.4$ to $0.6$) assigned across the training data, rather than derived from KDE-based density estimation. 
Table~\ref{tab:amp_results1} compares the performance of density-based $\sigma$ assignment with that of uniformly assigned $\sigma$ values.

Under the ID and ED metrics, the uniform-$\sigma$ variant performs slightly better than our density-based method; however, under the \emph{Performance} and \emph{H-Mean} metrics, our proposed density-based approach clearly outperforms it. A higher H-Mean indicates a better balance between diversity and quality of the generated samples.

It is also worth noting that the performance of the uniform-$\sigma$ model resembles that of dWJS with a fixed $\sigma = 0.5$ in terms of ID and ED. This is expected, since the average of the uniformly sampled $\sigma$ values is close to $0.5$. However, compared with fixed-$\sigma$ dWJS, the uniform-$\sigma$ model achieves better under the \emph{Performance} and \emph{H-Mean} metrics.

These observations suggest that using a \emph{range} of $\sigma$ values, even when sampled uniformly acts as a useful regularizer. Nonetheless, our density-based $\sigma$ estimation further improves both quality and diversity, highlighting the importance of aligning $\sigma$ values with the underlying data geometry.

\begin{table}[h!]
\centering
\caption{Comparison of model performance between desnity based $\sigma$ estimation and uniformly assigned $\sigma$ values  }
\label{tab:amp_results1}
\begin{tabular}{lcccc}
\toprule
\textbf{Method} & \textbf{H-Mean($\uparrow$)} & \textbf{Performance ($\uparrow$)} & \textbf{ID ($\uparrow$)} & \textbf{ED ($\uparrow$)} \\
\midrule
\textbf{DDS} & \textbf{0.73 ± 0.08} & 0.92 ± 0.10 & 45.93 ± 5.40 & 36.80 ± 6.41 \\
\textbf{Uniform } & 0.68 +0.11 & 0.75 ± 0.22  & \textbf{48.56 ± 3.86} & \textbf{40.82 ± 4.81} \\
\bottomrule
\end{tabular}
\end{table}

\subsection*{DDS produces structurally plausible sequences while avoiding the extreme memorization observed at high smoothing levels.}
While the biophysical property distributions 
reported in previous sections provide initial evidence that generated sequences resemble
natural proteins, these aggregate statistics may not fully capture whether individual
sequences are biologically viable. To address this concern, we performed three
complementary computational validation analyses that assess structural foldability,
sequence naturalness, and evolutionary plausibility.

\subsubsection*{DDS generates protein with high structural confidence}
\label{sec:structure_prediction}

A fundamental requirement for a functional protein is its ability to fold into a
stable three-dimensional structure. We evaluated the structural plausibility of
generated sequences using ESM-2~\cite{lin2023evolutionary}, a state-of-the-art protein
structure prediction model. For each generated sequence, we computed the predicted
Local Distance Difference Test (pLDDT) score, which quantifies the model's confidence
in its structural prediction on a scale of 0--100. Scores above 70 indicate confident
predictions, while scores above 90 suggest very high confidence in the predicted structure.

We predicted structures for 500 sequences from each model and report the mean pLDDT
scores in Table~\ref{tab:structure_confidence}. As a reference baseline, we also
computed pLDDT scores for held-out natural sequences from the validation set.

\begin{table}[ht]
\centering
\caption{Structure prediction confidence scores. Higher pLDDT indicates greater
confidence that the sequence can fold into a stable 3D structure. pLDDT $>70$
reports the number of sequences with confident structural predictions.}
\label{tab:structure_confidence}
\begin{tabular}{lcc}
\toprule
\textbf{Model} & \textbf{Mean pLDDT ($\uparrow$)} & \textbf{pLDDT $>70$ (count) ($\uparrow$)} \\
\midrule
Natural (reference) & 75.30 $\pm$ 0.013 & 500 \\
DDS (Ours) & 75.30 $\pm$ 0.015 & 498 \\
dWJS ($\sigma=0.5$) & 72.19 $\pm$ 0.036 & 396 \\
dWJS ($\sigma=1.0$) & 76.52 $\pm$ 0.011 & 500 \\
\bottomrule
\end{tabular}
\end{table}

\subsubsection*{DDS generates natural-looking protein sequences and avoids trivial memorization of the training data}
\label{sec:naturalness}

Protein language models trained on millions of evolutionary sequences capture complex
patterns of amino acid co-occurrence and positional dependencies that characterize
natural proteins. We leveraged ESM-2~\cite{lin2023evolutionary}, a large-scale protein
language model with 650 million parameters, to assess how ``natural'' our generated
sequences appear.

For each sequence, we computed the pseudo-perplexity by masking each position in turn
and measuring the model's ability to predict the true amino acid. Formally, for a
sequence $\mathbf{x} = (x_1, \ldots, x_L)$, the pseudo-perplexity is defined as:
\begin{equation}
\text{PPL}(\mathbf{x}) =
\exp\left( -\frac{1}{L} \sum_{i=1}^{L}
\log P(x_i \mid \mathbf{x}_{\setminus i}) \right),
\end{equation}
where $P(x_i \mid \mathbf{x}_{\setminus i})$ is the probability assigned by ESM-2 to the
true amino acid $x_i$ given all other positions. Lower pseudo-perplexity indicates that
the sequence is more consistent with the patterns learned from natural proteins.

\begin{table}[ht]
\centering
\caption{Evaluation of generated and real protein sequences using ESM-2
pseudo-perplexity (PPL) and mean log-probability. Lower PPL indicates higher
evolutionary plausibility, while a higher mean log-probability indicates better
alignment with the language model. Results are reported over the first 500 sequences.}
\label{tab:esm_ppl}
\begin{tabular}{lcc}
\toprule
\textbf{Model / $\sigma$} & \textbf{Pseudo-PPL $\downarrow$} & \textbf{Mean Log-Prob $\uparrow$} \\
\midrule
Natural (reference) & $2.51 \pm 0.46$ & $-0.90 \pm 0.17$ \\
DDS (Ours) & $2.88 \pm 0.39$ & $-1.05 \pm 0.14$ \\
dWJS ($\sigma = 0.5$) & $3.83 \pm 1.01$ & $-1.31 \pm 0.24$ \\
dWJS ($\sigma = 1.0$) & \textbf{$2.42 \pm 0.42$} & \textbf{$-0.87 \pm 0.17$} \\
\bottomrule
\end{tabular}
\end{table}

Taken together, Tables~\ref{tab:structure_confidence} and~\ref{tab:esm_ppl} suggest
that dWJS ($\sigma=1.0$) achieves the highest structure confidence and sequence
naturalness, while dWJS ($\sigma=0.5$) yields the lowest scores under both metrics.
However, these metrics alone cannot distinguish valid generalization from trivial
memorization of the training data, motivating an explicit analysis of sequence
homology.

\subsubsection*{DDS Balances Structural Validity with Evolutionary Novelty}
\label{sec:homology}

To assess whether generated sequences represent genuine evolutionary novelty rather
than memorized training examples, we performed sequence similarity searches using
BLAST~\citep{altschul1990basic} against the POAS training set. For each generated
antibody, homology was evaluated at the pair level by defining the sequence identity
as the minimum of the best BLAST identities of the heavy and light chains. This
conservative definition reflects the fact that a paired antibody is only as novel as
its less homologous chain.

We report the distribution of pair-level sequence identity in disjoint bins:
$<30\%$, 30--50\%, 50--70\%, 70--90\%, 90--95\%, and $>95\%$. The 70--90\% range
corresponds to sequences that are novel yet evolutionarily plausible, while identities
above 95\% indicate near-duplicate sequences and potential memorization.

\begin{table}[ht]
\centering
\caption{Sequence homology analysis against the POAS training set. Pair-level identity
is computed as the minimum of the best BLAST identities of the heavy and light chains.
Values are reported as percentages and sum to 100\%. Samples from all the models achieve a hit rate of 100\%}
\label{tab:homology}
\begin{tabular}{lcccccc}
\toprule
\textbf{Model / $\sigma$} &
\textbf{$<$30} &
\textbf{30--50} &
\textbf{50--70} &
\textbf{70--90} &
\textbf{90--95} &
\textbf{$>$95} \\
\midrule
DDS (Ours) & 0.0 & 0.0 & 0.0 & 38.6 & 36.4 & 25.0 \\
dWJS ($\sigma=1.0$) & 0.0 & 0.0 & 0.0 & 10.0 & 24.2 & 65.8 \\
dWJS ($\sigma=0.5$) & 0.0 & 1.2 & 14.2 & 76.6 & 6.4 & 1.6 \\
\bottomrule
\end{tabular}
\end{table}

The homology analysis reveals that the strong performance of dWJS ($\sigma=1.0$) on
structure and language-model metrics is largely driven by memorization, with 65.8\%
of generated sequences exceeding 95\% identity to the training set. In contrast,
dWJS ($\sigma=0.5$) substantially reduces memorization and concentrates the majority
of samples (76.6\%) in the desirable 70--90\% identity range, albeit at the cost of
reduced structural confidence. DDS occupies an intermediate regime, maintaining high
structural validity while exhibiting increased redundancy due to data-dependent
smoothing.


Together, these results highlight a fundamental trade-off between validity and
novelty. While structure prediction confidence and language-model likelihood favor
models that remain close to the training distribution, BLAST-based homology analysis
is essential for revealing memorization effects that are otherwise obscured. DDS
achieves a balanced trade-off, producing structurally plausible sequences while
avoiding the extreme memorization observed at high smoothing levels.

These results reveal that high apparent sample quality under large fixed smoothing can be driven by silent memorization, a failure mode that is not detectable from structural or language-model scores alone but is mitigated by density-aware smoothing.

\subsection*{DDS Generates Diverse Samples Matching the Ground Truth Immunogenicity Distributional Similarity} 


We also evaluate our method on a real-world antibody design task, focusing on whether DDS can generate diverse VH–VL sequences while preserving the immunogenicity distributional characteristics of the ground-truth data. We construct our training dataset from the \textbf{Coronavirus Antibody Database (CoV-AbDab)}\citep{raybould2021cov}, a curated repository of experimentally characterized antibodies binding to coronaviruses. CoV-AbDab provides antibody heavy and light chain variable-region sequences, along with metadata. The CoV-AbDab summary table was downloaded from the official repository\footnote{\url{https://opig.stats.ox.ac.uk/webapps/covabdab/}} as a CSV file containing antibody sequence annotations. Each entry includes variable heavy (VH) and variable light (VL) chain sequences.
We apply the a few  preprocessing steps to obtain a clean paired VH--VL dataset. We make sure to retain only classical antibodies by filtering entries labeled as \texttt{Ab}, excluding nanobodies (\texttt{Nb}).  Extract variable-region sequences from the \texttt{VHorVHH} (heavy) and \texttt{VL} (light) fields. 

After filtering, each remaining record contains a paired heavy and light chain. Duplicate VH--VL pairs are removed to avoid redundancy. The resulting dataset consists of paired variable-region sequences used for training in total of 11K samples.

\paragraph{Immunogenicity Evaluation}

We assessed immunogenicity risk using the \textbf{NetMHCIIpan} \emph{web server} provided by DTU Health Tech, rather than a local installation. For each antibody variable-region sequence, we generated overlapping 15-mer peptides and submitted them to the server in EL (eluted ligand) prediction mode. We retrieved the results and summarized peptide-level predictions into a sequence-level immunogenicity score by aggregating binding ranks across all peptides of a sequence.





\begin{table}[t]
\centering
\caption{Immunogenicity statistics and distribution matching. Lower is better for KS/WD ($\downarrow$).}
\label{tab:Cov_eval_immuno}
\begin{tabular}{lcccc}
\toprule
\textbf{Model} 
& \textbf{KS} $\downarrow$
& \textbf{WD} $\downarrow$
& \textbf{Mean score}
& \textbf{Max score} \\
\midrule
Real (train) 
& -- 
& -- 
& 0.0162 $\pm$ 0.0123
& 0.0756 \\

Ours     
& \textbf{0.1688 $\pm$ 0.0176}
& 0.0043 $\pm$ 0.0006
& 0.0149 $\pm$ 0.0136
& 0.0656 \\

dWJS($\sigma = 0.5$)
& 0.1804 $\pm$ 0.0180
& 0.0041 $\pm$ 0.0006
& 0.0169 $\pm$ 0.0141
& 0.0647 \\

dWJS($\sigma = 1.0$)
& 0.2868 $\pm$ 0.0167
& \textbf{0.0040 $\pm$ 0.0005}
& 0.0191 $\pm$ 0.0119
& 0.0604 \\
\bottomrule
\end{tabular}
\end{table}

\begin{table}[t]
\centering
\caption{Novelty and diversity of generated antibodies. Higher is better for ED/ID ($\uparrow$). Both ED and ID are normalised.}
\label{tab:Cov_eval_novelty}
\begin{tabular}{lcc}
\toprule
\textbf{Model} 
& \textbf{ED} $\uparrow$
& \textbf{ID} $\uparrow$ \\
\midrule

Ours     
& 0.334 $\pm$ 0.0780
& \textbf{0.507 $\pm$ 0.0040} \\

dWJS($\sigma = 0.5$)
& \textbf{0.344 $\pm$ 0.686}
& 0.484 $\pm$ 0.048 \\

dWJS($\sigma = 1.0$)
& 0.155 $\pm$ 0.0505
& 0.312 $\pm$ 0.0056 \\
\bottomrule
\end{tabular}
\end{table}

We compare the distribution of immunogenicity scores between generated and real antibodies using two distributional similarity metrics, KS and Wasserstein.

Table~\ref{tab:Cov_eval_immuno} and Table~\ref{tab:Cov_eval_novelty} summarize immunogenicity distribution matching and novelty/diversity computed on concatenated VH--VL sequences. Our model achieves the closest alignment to the real immunogenicity distribution, obtaining the lowest KS statistic, while also maintaining strong novelty and the highest intra-diversity, suggesting it explores the sequence space without evidence of memorization. In contrast, dWJS ($\sigma{=}1.0$) yields sequences that are substantially less novel and less diverse, consistent with partial memorization of training examples. dWJS ($\sigma{=}0.5$) attains comparable novelty to our model, but with a modest drop in intra-diversity, indicating slightly reduced coverage of the sequence space. In addition to KS/WD, Table~\ref{tab:Cov_eval_immuno} reports the mean and maximum immunogenicity scores, which summarize the typical and worst-case (tail) sequence-level risk after aggregating peptide-level binding ranks. Our model achieves a slightly lower mean than the real set and a lower max than the real-set maximum, indicating that DDS matches the distribution while avoiding rare high-risk outliers.

\subsection*{Selection of $\sigma$ range is important for DDS performance:}

The choice of the smoothing range, defined by $\sigma_{\min}$ and $\sigma_{\max}$, plays a crucial role in the performance of DDS. To study the sensitivity of the model to these hyperparameters, we conducted an ablation analysis on the AMP dataset, as shown in Table~\ref{tab:sigma_ablation}. In this study, $\sigma_{\min}$ was varied from 0.3 to 0.9, and $\sigma_{\max}$ from 0.5 to 1.0, considering all valid pairs with $\sigma_{\max} > \sigma_{\min}$. For each pair, we evaluated multiple $\sigma$ bandwidths ranging from 0.1 to 0.6.

The results indicate that the ranges $0.4$--$0.6$ and $0.4$--$0.7$ yield consistently strong performance compared to other settings. For our AMP experiments, we selected the range $0.4$--$0.6$. Although the ED metric is slightly higher for this $0.4$--$0.7$  range, the overall quality, as reflected by the performance score, is better for $0.4$--$0.6$, $92$ whereas for the range $0.4$--$0.7$ is $0.87$. 

For the antibody sequence generation task, we reused the same range ($0.4$--$0.6$). We did not run an exhaustive ablation for antibodies due to computational constraints; however, we observed that the sigma ranges performing well for AMP also transfer effectively to the other datasets.
\begin{table}[h!]
\centering
\scriptsize
\setlength{\tabcolsep}{2pt}
\caption{Ablation over $(\sigma_{\text{low}}, \sigma_{\text{high}})$ for AMP dataset. Each cell shows \textbf{HM / ED}.}
\label{tab:sigma_ablation}
\begin{tabular}{c|cccccc}
\toprule
$\sigma_{\text{low}} \backslash \sigma_{\text{high}}$ & 0.5 & 0.6 & 0.7 & 0.8 & 0.9 & 1.0 \\
\midrule
0.3 &   0.61 / 43.49 & 0.68 / 40.67 & 0.72 / 36.21 & 0.63 / 29.27 & 0.51 / 22.22 & 0.51 / 22.22 \\
0.4 &  0.65 / 42.10 & \textbf{0.73 / 36.80} & \textbf{0.73 / 38.64} & 0.64 / 29.45 & 0.48 / 20.08 & 0.30 / 10.65 \\
0.5 &  -- & 0.72 / 35.51 & 0.68 / 32.20 & 0.57 / 24.98 & 0.49 / 20.70 & 0.31 / 11.08 \\
0.6 &  -- & -- & 0.68 / 31.66 & 0.51 / 21.29 & 0.51 / 21.29 & 0.29 / 10.49 \\
0.7 &  -- & -- & -- & 0.60 / 26.72 & 0.46 / 18.98 & 0.32 / 11.68 \\
0.8 &  -- & -- & -- & -- & 0.35 / 12.98 & 0.29 / 10.50 \\
0.9 &  -- & -- & -- & -- & -- & 0.28 / 9.73 \\
\bottomrule
\end{tabular}
\end{table}
\section*{Methodological Background}

Neural Empirical Bayes (NEB)~\citep{saremi2019neural} provides a principled framework that combines non-parametric smoothing with empirical Bayes estimation to perform least–squares denoising. Given a clean sample $x$, a noisy counterpart is produced by adding Gaussian perturbation with variance $\sigma^2$:
\begin{equation}
    y = x + \mathcal{N}(0, \sigma^2).
\end{equation}
The empirical Bayes estimate of the clean signal is then given by
\begin{equation}
    \bar{x} = y + \sigma^2 \nabla \log f(y),
\end{equation}
where $f(y)$ is the density of the smoothed distribution. Although $f(y)$ is unknown, its score $\nabla \log f(y)$ can be approximated by a neural network.

The network is trained through a denoising objective that encourages accurate reconstruction of clean samples from their noisy versions:
\begin{equation}
    L(\phi) = \mathbb{E}_{x \sim p(x),\, y \sim p(y|x)}
    \left\| x - \hat{x}_\phi(y) \right\|^2,
\end{equation}
where $p(y|x) = \mathcal{N}(x, \sigma^2)$ and 
\[
    \hat{x}_\phi(y) = y + \sigma^2 g_\phi(y),
\]
with $g_\phi(y)$ acting as a learned approximation of the score function $\nabla \log f(y)$.

The Discrete Walk--Jump Sampling (dWJS) method~\citep{frey2023protein} adapts NEB to protein and antibody sequence generation. Each amino acid sequence is represented as a concatenated one-hot encoding: for a sequence of length $297$, this yields a binary vector of dimension $297 \times 21$. Gaussian noise is added to this continuous relaxation, allowing the model to learn the local structure of the data manifold.

After training the score model, sampling proceeds in two phases. A random sequence is first converted into its one-hot representation and perturbed with Gaussian noise, producing a continuous embedding. The sample is then iteratively updated by ascending the learned score field the ``walk'' step—thereby moving along the estimated energy landscape. After $T$ such refinement steps, the updated embedding is projected back to the discrete domain by selecting the most probable amino acid at each position, producing a valid protein sequence. This projection forms the ``jump'' step.

\paragraph{Sampling with a $\sigma$-conditioned score model}

\textbf{Input.} 
A $\sigma$-conditioned score network $g_{\phi}(y,\sigma)$; empirical noise-level distribution $\hat{p}(\sigma)$; step size $\delta$; number of iterations $T$.

\textbf{Output.} 
A sample $\hat{x}$ from the learned data distribution.

\textbf{Procedure.}

\begin{enumerate}
    \item \textbf{Sample a noise scale}
    \[
    \sigma \sim \hat{p}(\sigma).
    \]

    \item \textbf{Initialize the latent variable}
    \[
    y_0 \sim \mathcal{N}(0,\sigma^2 I_d).
    \]

    \item \textbf{Iterative Langevin updates.}  
    For $t = 0, \ldots, T-1$, update $y_t$ according to
    \[
    y_{t+1} = y_t + \delta\, g_{\phi}(y_t,\sigma)
    + \sqrt{2\delta}\,\varepsilon_t,
    \qquad
    \varepsilon_t \sim \mathcal{N}(0,I_d).
    \]

    \item \textbf{Denoising step.}
    \[
    \hat{x} = y_T + \sigma^2 g_{\phi}(y_T,\sigma).
    \]
\end{enumerate}

\subsection*{Relation to Geometric Guidance}
Our framework can be theoretically linked with Classifier-Free Guidance (CFG) \citep{ho2021classifier}, albeit with two critical distinctions that tailor it to unsupervised manifold learning.

The standard CFG update modifies the score estimate $\tilde{\epsilon}$ by extrapolating between an unconditional score $\epsilon_\theta(y_t)$ and a conditional score $\epsilon_\theta(y_t, c)$ (where $c$ is a class label):
\begin{equation}
    \tilde{\epsilon}_\theta(y_t, c) = \epsilon_\theta(y_t) + w \cdot (\epsilon_\theta(y_t, c) - \epsilon_\theta(y_t))
\end{equation}
where $w$ is the guidance scale. Typically, $w > 1$ is used to amplify conditional probability at the expense of diversity.

Our Density-Dependent Smoothing (DDS) replaces the discrete class label $c$ with the continuous noise scale $\sigma(x)$, which serves as a proxy for the local geometric density. Our sampling procedure utilizes the conditional score model $g_\phi(y_t, \sigma)$ directly. This is mathematically equivalent to the CFG formulation where the guidance weight is fixed at \textbf{$w=1$} (standard conditional sampling):
\begin{equation}
    \tilde{g}_\phi(y_t) = g_\phi(y_t, \sigma)
\end{equation}
This choice ensures that the model faithfully respects the learned geometry without introducing the mode-collapse artifacts often associated with high-guidance sampling ($w > 1$).

Furthermore, we extend the notion of the ``conditioning class'' from a discrete set to a \textbf{continuous spectrum}. While standard CFG conditions on discrete labels $c \in \{1, \dots, K\}$, our model conditions on the continuous noise scale $\sigma \in \mathbb{R}^+$. By treating $\sigma$ as a proxy for local density, we effectively define an \textbf{arbitrary number of geometric classes}, ranging from ``highly dense'' to ``sparse.'' This allows the score function to adapt smoothly across the heterogeneous energy landscape of protein sequences. 

\paragraph{Evaluation Metrics}}
To assess the quality of generated antibody sequences, we evaluate a diverse set of metrics that capture uniqueness, diversity, novelty, edit distance from held-out data, and distributional similarity.
\begin{itemize}

\item \textbf{Uniqueness (U).}  
Uniqueness measures the fraction of generated sequences that are distinct from one another.  
A higher value indicates that the model avoids producing redundant or repeated samples.  
Formally, if $\mathcal{D}$ is the set of generated sequences and $\text{Unique}(\mathcal{D})$ denotes the number of distinct elements, then
\[
U = \frac{\text{Unique}(\mathcal{D})}{|\mathcal{D}|}.
\]

\item \textbf{Intra-Diversity (ID).}  
Intra-diversity quantifies how different the generated samples are from each other.  
Diverse sets of sequences are desirable because they explore a broader region of the sequence space.  
Following prior work, we compute diversity using the average pairwise Levenshtein distance:
\[
\text{ID}(\mathcal{D}) = 
\frac{\sum_{x_i \in \mathcal{D}} \sum_{x_j \in \mathcal{D} \setminus \{x_i\}} d(x_i, x_j)}
      {|\mathcal{D}| (|\mathcal{D}| - 1)},
\]
where $d(\cdot,\cdot)$ is the Levenshtein distance.  
Larger values correspond to greater diversity among generated sequences.

\item \textbf{Edit Distance (ED).}  
Edit Distance measures how different the generated sequences are from the samples from validation set $\mathcal{D}_0$.  
This metric evaluates whether the model generates sequences that are novel and not present in the known dataset.  
We compute ED as the average minimum distance from each generated sequence to any sequence in $\mathcal{D}_0$:
\[
\text{ED}(\mathcal{D}) =
\frac{1}{|\mathcal{D}|}
\sum_{x_i \in \mathcal{D}}
\min_{s_j \in \mathcal{D}_0} d(x_i, s_j).
\]
Higher ED values indicate stronger novelty.

\item \textbf{DCS (Distributional Consistency Score).}  
DCS~\citep{frey2023protein} is analogous to the Fréchet Inception Distance (FID) for images.  
It evaluates how consistent the generated sequences are with the reference distribution while still maintaining novelty and diversity.  
Higher DCS values indicate better alignment with the true distribution.

\item \textbf{Wasserstein Distance (WD).}  
Wasserstein distance measures how closely the generated sequences match the biochemical property distribution of the validation set.  
For each sequence, we compute biochemical properties using \texttt{BioPython}~\citep{cock2009biopython} (e.g., instability index, aromaticity).  
We then compute the average Wasserstein distance between the property distributions of the generated and validation sets:
\[
\text{WD} = W(\mathcal{P}_{\text{gen}}, \mathcal{P}_{\text{val}}).
\]
Lower WD indicates that the generated sequences exhibit realistic biochemical characteristics.
\item \textbf{KS-Statistic}
It measures the maximum difference between empirical cumulative distribution functions:
\[
\text{KS} = \max_x \left| F_{\text{real}}(x) - F_{\text{gen}}(x) \right|,
\]

where $F_{\text{real}}$ and $F_{\text{gen}}$ denote cumulative distributions of real and generated immunogenicity scores, respectively.

\end{itemize}

\section*{Conclusion}

More broadly, our findings indicate that respecting the heterogeneous geometry of biological sequence space is not an implementation detail but a prerequisite for reliable generative modeling. Fixed global smoothing assumptions—common across current diffusion- and language-model-based generators—systematically distort sparse biological manifolds, leading to either oversmoothing of dense functional clusters or fragmentation in sparse regions. Geometry-aware generative mechanisms such as \textbf{Density-Dependent Smoothing (DDS)} provide a principled alternative, enabling scalable and trustworthy molecular design in regimes that are fundamentally inaccessible to fixed-noise models.


Across a controlled multimodal toy benchmark, DDS improves fidelity while maintaining exploration, recovering true (including held-out) modes with substantially fewer spurious generations. On real biological design tasks, large-scale antibody repertoire generation, CDR mutant generation, antimicrobial peptide design, and coronavirus antibody generation, DDS consistently yields a stronger quality and diversity trade-off than fixed-$\sigma$ baselines. Importantly, complementary evaluations using structure prediction and protein language-model scores, together with BLAST-based~\citep{altschul1990basic} homology analysis, show that high validity under fixed large smoothing can be driven by memorization, whereas DDS better balances structural plausibility with evolutionary novelty. Finally, on CoV-AbDab, DDS matches the ground-truth immunogenicity distribution more closely while maintaining strong novelty/diversity, suggesting that density-aware noise scaling can improve distributional faithfulness in safety-relevant properties.

We anticipate that geometry-aware smoothing will serve as a foundational design principle for future generative models of biological molecules. By aligning stochastic smoothing with the intrinsic geometry of sparse and heterogeneous fitness landscapes, approaches such as DDS offer a path toward scalable, reliable, and safety-conscious molecular design beyond the limits imposed by fixed-noise generative frameworks.

\section*{Limitations and Future Work}

While DDS provides a simple and effective method, our approach also has some limitations that motivate future research. 
Our method relies on selecting a predefined range for scaling the per-sample $\sigma$ values. The choice of lower and upper bounds can influence model performance. 
Developing a more principled or adaptive mechanism for setting these bounds remains an open problem. 

Our current study is primarily empirical. While the results on synthetic datasets and biomolecular sequence design tasks strongly suggest that data-adaptive $\sigma$ improves the quality of generated samples, a theoretical understanding of why and when adaptive noise scaling enables better approximation of the underlying score function is still lacking. 
We aim to extend this work with a formal analysis, potentially establishing error bounds that connect local density estimates, $\sigma$ selection, and the accuracy of score approximation.  

\bibliography{sn-bibliography}

\newpage
\section*{Appendices}
\begin{appendices}

\begin{figure}[b]
  \centering
  \begin{subfigure}[b]{0.47\linewidth}
    \centering
    \includegraphics[width=\linewidth]{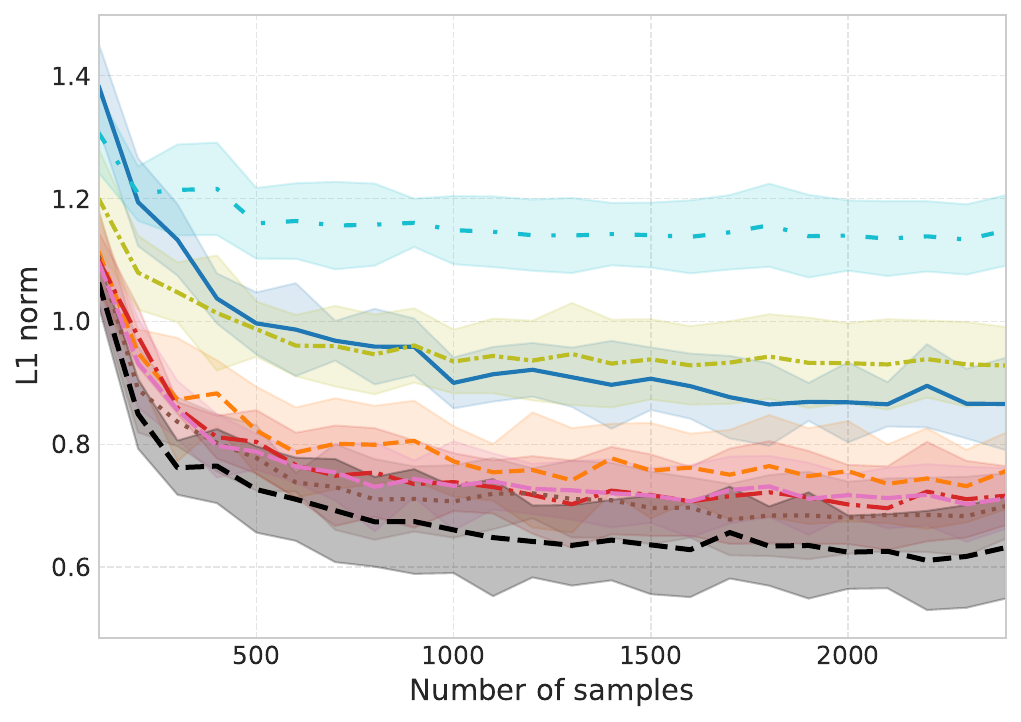}
    \caption{$L_1$ norm}
    \label{fig:l1}
  \end{subfigure}\hfill
  \begin{subfigure}[b]{0.47\linewidth}
    \centering
    \includegraphics[width=\linewidth]{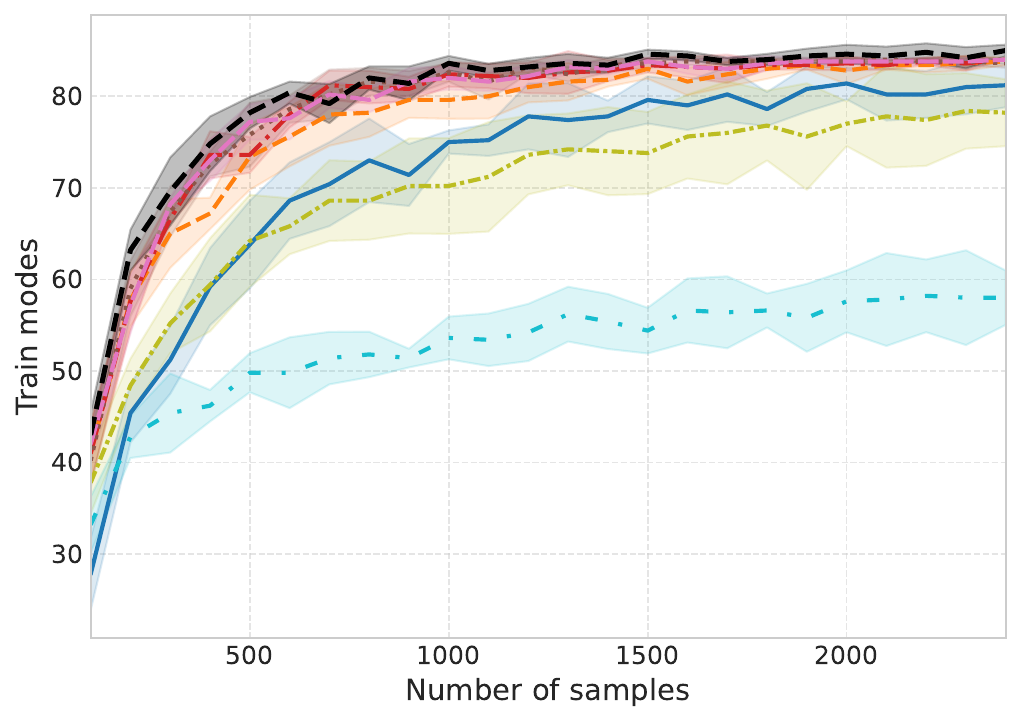}
    \caption{True modes covered}
    \label{fig:true}
  \end{subfigure}\hfill
  \begin{subfigure}[b]{0.47\linewidth}
    \centering
    \includegraphics[width=\linewidth]{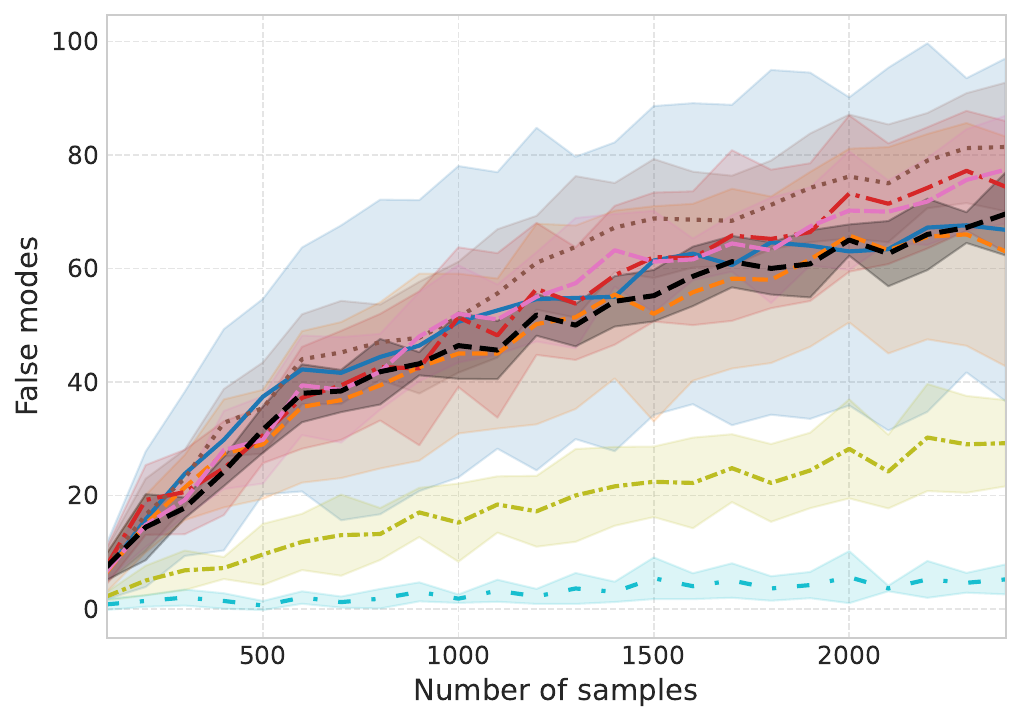}
    \caption{False modes covered}
    \label{fig:false}
  \end{subfigure}\hfill
    \begin{subfigure}[b]{0.47\linewidth}
    \centering
    \includegraphics[width=\linewidth]{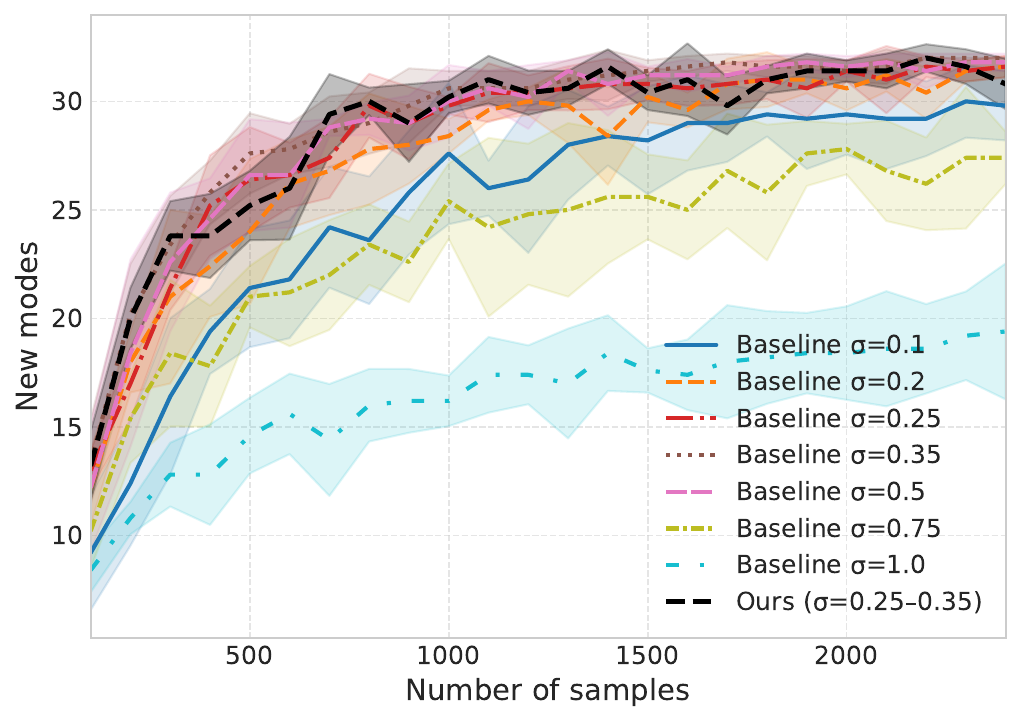}
    \caption{New modes covered}
    \label{fig:false}
  \end{subfigure}

  \caption{Performance on the synthetic 4D dataset}
  \label{fig:synthetic_modes_full}
\end{figure}

\paragraph{Extended baseline sweep on the toy dataset.}
In the main text, we compare against a reduced set of $\sigma$ values to keep Figure~\ref{fig:synthetic_modes} readable; here we in Figure~\ref{fig:synthetic_modes_full} we include the complete baseline sweep over fixed smoothing scales $\sigma \in \{0.1, 0.2, 0.25, 0.35, 0.5, 0.75, 1.0\}$. This extended view confirms that baseline performance is sensitive to the choice of smoothing scale: very small $\sigma$ can over-explore and introduce spurious modes, whereas overly large $\sigma$ over-smooths and under-covers true modes. DDS remains consistently competitive across the sweep, avoiding both failure modes and reducing the need for precise $\sigma$ tuning.

\end{appendices}



\end{document}